\documentstyle[eqsecnum,aps]{revtex}
\begin{document}
\draft
\title{Nonasymptotic effects in critical sound propagation\\
associated with spin-lattice relaxation}
\author{A. Pawlak}
\address{Institute of Physics, A. Mickiewicz University, Pozna\'{n}, Poland}
%\date{\today }
\maketitle

\begin{abstract}
The nonasymptotic critical behavior of sound attenuation coefficient has
been studied in an elastically isotropic Ising system above the critical
point on the basis of a complete stochastic model including both spin-energy
and lattice-energy modes linearly coupled to the longitudinal sound mode.
The effect of spin-lattice relaxation on the ultrasonic attenuation is
investigated. The crossover between weak-singularity behavior $t^{-2\alpha }$
and strong-singularity behavior $t^{-(z\nu +\alpha )}$ is studied as
dependent on the values of ultrasonic frequency, reduced temperature,
relaxation times etc.. A new high-frequency regime with a singularity of the
type $t^{-z\nu +\alpha }$ is discovered in the magnetic systems. This new
regime corresponds to an adiabatic sound propagation and is very similar to
the ones in binary mixture and liquid helium. The scaling functions are
given in all regimes to the first order in $\epsilon $. A new
frequency-dependent specific-heat being the harmonic average of the bare
lattice and critical spin specific-heats is introduced. It was shown that
such specific-heat describes the process of equilibrization between spin and
lattice subsystems and includes the most important features of critical
sound attenuation. In some regions of coupling constants the acoustic
self-energy can be very well approximated solely by this quantity.
\end{abstract}


\section{INTRODUCTION}

A strong anomaly of sound attenuation measured in Ising like magnetic metals 
\cite{lut,kaw76} is well described by the theories \cite{mur,iros,foss,paw89}
in which the sound wave is assumed to couple to two critical spin
fluctuations. The sound attenuation coefficient is found then as the
imaginary part of the four-spin response function. On the contrary to
metals, in magnetic insulators we observe a weak anomaly in sound
attenuation \cite{lut,kaw76}, which has been phenomenologically explained 
\cite{kaw69} by postulating a linear coupling of the sound mode to the
spin-energy with the spin-lattice (or more correctly
spin-energy-lattice-energy) relaxation time playing the essential role.
Assuming only the specific-heat like singularity of the spin-lattice
relaxation time and simple separability of contributions coming from both
types of coupling, Kawasaki \cite{kaw69} was able to obtain weak divergence
of the sound attenuation coefficient proportional to the square of the
specific-heat for the magnets where the first coupling is negligible.
However, as it was shown by the renormalization-group analysis of the model
of coupled spin and energy fields \cite{hhm76} the first assumption is not
correct sufficiently close to the critical point where the energy relaxes
with the same characteristic exponent as the spin fluctuations, $z_E=z$,
because the singular part of the energy response function is dominated by
the four-spin response function. Also the second assumption concerning the
separability fails near $T_C$ as was recently shown \cite{pawepj}. The
objective of this paper is to use a complete stochastic model which includes
interactions between the order parameter (spin), the acoustic phonon,
spin-energy and lattice-energy modes in order to find the asymptotic as well
as nonasymptotic behavior of the sound attenuation coefficient. We shall
restrict our discussion to the disordered phase. We consider an elastically
isotropic Ising model, where above the critical temperature a phonon
(longitudinal) mode $Q$ is coupled to a scalar order parameter $S$ as well
as to the fluctuations of spin- and lattice- energy densities $e_S$ and $e_L$%
. Introduction of the second energy density associated with the lattice (or
conduction electrons in some metals) allows us to consider much richer
dynamics than that of the two limiting cases: pure relaxation to a reservoir
with infinite thermal conductivity or pure diffusion of energy. Our model
permits also a study of the intermediate dynamics between these limiting
cases. We have obtained an general expression for the sound attenuation
coefficient properly describing the ultrasound attenuation in very broad
temperature and frequency range. The purpose of this paper is to demonstrate
that the strong sound attenuation anomaly (sometimes called
Murata-Iro-Schwabl behavior) as well as the weak anomaly (Kawasaki behavior)
can be obtained within our stochastic model depending on the relative size
of the reduced temperature, frequency and other parameters of the model. The
nonuniversal amplitudes for both types of behavior are modified considerably
by the presence of the second energy field in comparison with
one-energy-field model. In the high-frequency regime we obtain a new kind of
critical behavior, which we shall call the ``adiabatic'' behavior, analogous
to the critical attenuation behavior in $^4\text{He}$ and in binary
mixtures. Contrary to the Murata-Iro-Schwabl and Kawasaki behavior the
amplitude for the ``adiabatic'' limit is determined completely only by one
coupling constant, the one describing the interaction of the sound mode with
two fluctuations of the order parameter. The ``adiabatic'' limit does not
change by the inclusion of the field $e_L$. We shall also show that in some
regions of parameters the acoustic self-energy is simply proportional to a
frequency-dependent specific-heat $C_{-}$ which is the harmonic average of
the bare lattice specific-heat and the specific-heat of an idealized spin
system (model A in terminology of \cite{hhm74}). $C_{-}$ has a simple
interpretation : it equals to the ratio of the heat transferred from one
sub-system to the other, to the induced temperature difference between
subsystems. The important point is that such specific-heat shows three main
types of singularities characterizing the sound attenuation coefficient as
the critical temperature is approached for low or high frequencies.

The paper is organized as follows. In Sec.II we introduce the model and then
perform some dynamical decoupling transformations. In Sec.III we obtain a
general expression for the acoustic self-energy taking into account the
reducibility of the latter with respect to energy propagators. In Sec.IV
many different regimes for sound-attenuation are discussed as a function of
frequency, temperature and relaxation times. The scaling functions are
determined within the one-loop approximation. We then show that for some
regions of parameters the ultrasonic attenuation can be written as the
imaginary part of $C_{-}(\omega )$. The relevance of this specific-heat to
the description of nonasymptotic sound attenuation is shown. In Sec.V we
summarize the results.

\section{MODEL AND BASIC FORMULAS}

\subsection{Equations of motion}

The entropy of the system \cite{hhm74,paw97} can be written as 
\begin{eqnarray}
H &=&\frac 12\int d^d\!x\left\{ rS^2+(\nabla S)^2+\frac u2S^4+C_{12}\left(
\sum_\alpha e_{\alpha \alpha }\right) ^2+2C_{44}\sum_{\alpha ,\beta
}e_{\alpha \beta }^2\right.   \nonumber \\
&&\left. \mbox{}+2g\sum_\alpha e_{\alpha \alpha
}S^2+2fe_SS^2+2w(e_S+ae_L)\sum_\alpha e_{\alpha \alpha }+\frac{e_S^2}{C_S}+%
\frac{e_L^2}{C_L}\right\} \ ,  \label{ham}
\end{eqnarray}
where $e_{\alpha \beta }(x)$ are components of the strain tensor 
%and $g$, $w$%, $aw$, and $f$ are the bare coupling constants. 
and the symbols $C_{\alpha \beta }$ stand for the bare elastic constants and 
$C_S$ and $C_L$ are the spin and the lattice specific-heat, respectively.
The unitary mass density and $k_BT_C=1$ have been assumed. The first three
terms in the total Hamiltonian- describing the static behavior of the
system, make the Ginzburg-Landau functional for the spin variable. The
elastic energy is given by the 4th and 5th term in Eq.\ (\ref{ham}). Here,
we have made use of the relation $C_{11}-C_{12}=2C_{44}$ applicable to the
isotropic cubic systems. The last two terms results from lowest order
expansion of the entropy functional with respect to energy fields. The other
terms in the Hamiltonian describe interactions. The constant $g$ denotes the
coupling of the (longitudinal) sound mode to two spin fluctuations. The
coupling of the sound to the energy fields is characterized by the constants 
$w$ and $aw$ where the parameter $a$ is the ratio of the couplings of the
phonon mode to the lattice-energy $e_L$ and spin-energy density $e_S$,
respectively, $f$ is the coupling constant between the order parameter and
the spin energy generating the divergence of the specific-heat.

After introducing the normal mode expansion of the strain tensor, the
dynamics of the system can be described by the coupled Langevin equations 
\cite{paw89,paw97} 
\begin{mathletters}
\label{eqsmot}
\begin{equation}
\dot{S}_{{\bf k}}=-\Gamma \frac{\delta H}{\delta S_{{\bf -k}}}+\xi _{{\bf k}%
},
\end{equation}
\begin{equation}
\ddot{Q}_{{\bf k}}=-\frac{\delta H}{\delta Q_{{\bf -k}}}-\theta k^2\dot{Q}_{%
{\bf k}}+\eta _{{\bf k}},
\end{equation}
\begin{equation}
\dot{e}_{{\bf k}}^S=-(\gamma _S+\lambda _Sk^2)\frac{\delta H}{\delta e_{{\bf %
-k}}^S}+\gamma \frac{\delta H}{\delta e_{{\bf -k}}^L}+\varphi _{{\bf k}},
\label{2c}
\end{equation}
\begin{equation}
\dot{e}_{{\bf k}}^L=-(\gamma _L+\lambda _Sk^2)\frac{\delta H}{\delta e_{{\bf %
-k}}^L}+\gamma \frac{\delta H}{\delta e_{{\bf -k}}^S}+\psi _{{\bf k}},
\label{2d}
\end{equation}
where $Q_{{\bf k}}$ is the longitudinal phonon normal coordinate and $\xi _{%
{\bf k}},\eta _{{\bf k}},\varphi _{{\bf k}}$ and $\psi _{{\bf k}}$ are the
Fourier components of Gaussian white noises with variances related to the
bare damping terms $\Gamma ,\theta k^2,(\gamma _S+\lambda _Sk^2)$ and $%
(\gamma _L+\lambda _Lk^2)$ through the Einstein relations.

The first two of these equations have been commonly used in investigation of
critical sound propagation \cite{iros,paw89,deng,paw901}. The other two
describe the energy flows between the sub-systems. In the absence of the
nonlinear terms and noises they transform into the equations describing the
decay of total energy $e=e_S+e_L$ and equalization of temperatures of both
sub-systems 
\end{mathletters}
\begin{equation}
\dot{e}=-\frac{\Lambda _{+}}{C_{+}}e-\Lambda _{\Delta e}C_{-}\Delta 
\end{equation}
\begin{equation}
\dot{\Delta}=-\frac{\Lambda _{\Delta e}}{C_{+}}e-\Lambda C_{-}\Delta 
\end{equation}
where $\Delta =\delta T_L-\delta T_S=e_L/C_L-e_S/C_S$ is the temperature
difference between both sub-systems, $\Lambda _{+}=\gamma _S+\gamma
_L-2\gamma +\lambda _{+}k^2$, $\Lambda _{\Delta e}=\frac{\gamma _L-\gamma
+\lambda _Lk^2}{C_L}-\frac{\gamma _S-\gamma +\lambda _Sk^2}{C_S}$ , $\Lambda
=\frac{\gamma _S+\lambda _Sk^2}{C_S^2}+\frac{\gamma _L+\lambda _Lk^2}{C_L^2}+%
\frac{2\gamma }{C_SC_L}$, $\lambda _{+}=\lambda _S+\lambda _L,C_{+}=C_S+C_L,$
and $C_{-}=(1/C_S+1/C_L)^{-1}$ . If $\gamma _S=\gamma _L=\gamma $, then the
total energy is conserved (for $k\rightarrow 0$) and $\frac{\Lambda _{+}}{%
C_{+}}=\frac{\lambda _{+}}{C_{+}}k^2$ is the rate of the thermal conduction
with $\lambda _S$ and $\lambda _L$ being the thermal conductivity for spin
and lattice sub-systems, respectively. For this case, the temperature
difference $\Delta $ decays at the rate $\Lambda C_{-}\approx \gamma /C_{-}$%
. Therefore, $\Lambda C_{-}$ can be interpreted as the frequency of the
spin-lattice relaxation. It reduces to $\gamma /C_S$ for the lattice of
infinite specific heat. For non-conserved total energy, as e.g. for $\gamma
_L>\gamma _S=\gamma $ (we assume that the spin-energy can change only
through interactions with lattice degrees of freedom i.e. $\gamma _S=\gamma $
), 
\begin{equation}
\dot{e}=-\frac{\Lambda _{+}}{C_{+}}e-\frac{\Lambda _{+}C_{-}}{C_L}\Delta 
\end{equation}
\begin{equation}
\dot{\Delta}=-\frac{\Lambda _{+}}{C_{+}C_L}e-\Lambda C_{-}\Delta 
\end{equation}
with $\Lambda _{+}\simeq \gamma _L-\gamma $. The total energy, in the
absence of the temperature difference $\Delta $, decays at the frequency $%
(\gamma _L-\gamma )/C_{+}$ and the temperature difference relaxes (in the
absence of total energy fluctuations) at the rate $\Lambda C_{-}=\frac \gamma
{C_{-}}+\frac{\Lambda _{+}C_S}{C_{+}C_L}$ . 
%In the limit $\gamma _L\rightarrow
%\infty $ the lattice-energy always takes its equilibrium value and only the
%equation 
%\[
%\dot{e}_S=-\gamma _S\frac{\delta H}{\delta e_S},
%\]
%matters in our considerations. Thus we recover the model without
%lattice-energy field \cite{paw94,pawepj}.

In a typical case $e_L$ is the sum of energies of all phonon branches except
for the longitudinal acoustic phonons with wave-vectors restricted to the
sphere $0<|{\bf k|}\ll 1$. These phonons are taken explicitly into account
in the model as the quantity of our primary interest in this paper is the
phonon response function. In some metals, where the lattice can be treated
fully adiabatically ($C_L\rightarrow \infty $), $e_L$ may be understood as
the energy density of conduction electrons. In that case, the total energy
in the system composed of the localized electrons and the conduction
electrons sub-systems is no longer conserved so we have admitted the
possibility of nonconserved total energy $e_L+e_S$ ($\gamma _L\neq \gamma $)
in the equations of motion.

\subsection{Lagrangian}

Instead of using Eq.\ (\ref{eqsmot}), it is convenient to represent the
model in terms of the equivalent functional form \cite{jans,bausch} with a
Lagrangian given by 
\begin{eqnarray}
L &=&\int_\omega \sum_{{\bf k}}\left\{ \Gamma \tilde{S}_{{\bf k},\omega }%
\tilde{S}_{-{\bf k},-\omega }+\theta k^2\tilde{Q}_{{\bf k},\omega }\tilde{Q}%
_{-{\bf k},-\omega }+\Gamma _S(k)\tilde{e}_{{\bf k},\omega }^S\tilde{e}_{-%
{\bf k},-\omega }^S+\Gamma _L(k)\tilde{e}_{{\bf k},\omega }^L\tilde{e}_{-%
{\bf k},-\omega }^L-2\gamma \tilde{e}_{{\bf k},\omega }^S\tilde{e}_{-{\bf k}%
,-\omega }^L\right.   \nonumber \\
&&-\tilde{Q}_{{\bf k},\omega }\left[ (-\omega ^2+i\theta k^2\omega )Q_{-{\bf %
k},-\omega }+\frac{\partial H}{\partial Q_{{\bf k},\omega }}\right] -\tilde{S%
}_{{\bf k},\omega }\left( i\omega S_{-{\bf k},-\omega }+\Gamma \frac{%
\partial H}{\partial S_{{\bf k},\omega }}\right)   \nonumber \\
&&\left. -\tilde{e}_{{\bf k},\omega }^S\left[ i\omega e_{-{\bf k},-\omega
}^S+\Gamma _S(k)\frac{\partial H}{\partial e_{{\bf k},\omega }^S}-\gamma 
\frac{\partial H}{\partial e_{{\bf k},\omega }^L}\right] -\tilde{e}_{{\bf k}%
,\omega }^L\left[ i\omega e_{-{\bf k},-\omega }^L+\Gamma _L(k)\frac{\partial
H}{\partial e_{{\bf k},\omega }^L}-\gamma \frac{\partial H}{\partial e_{{\bf %
k},\omega }^S}\right] \right\} \ ,
\end{eqnarray}
where $\tilde{S},\tilde{Q},\tilde{e}_S$ and $\tilde{e}_L$ are auxiliary
``response'' fields and $\Gamma _i(k)=\gamma _i+\lambda _ik^2$ for $i=S,L$.
The Lagrangian does not contain the Jacobian because we have assumed the
Heaviside step function to be zero for time $t=0$. This assumption excludes
the accusal terms in the perturbation theory \cite{jans79}.%
%Although the variables $Q, e_S$ and $e_L$ might be integrated out
%straightforwardly from the statistical weight , it is more
%convenient for our purposes to use series of Gaussian transformations. First
%of them, 
In this formalism the correlation and response functions are given by path
integrals weighted with a density $exp(L)$. For instance if the phonon
variable is coupled to an external field $h$ in the entropy functional (\ref
{ham}), an additional contribution $h_{{\bf k,}\omega }\tilde{Q}_{-{\bf k,-}%
\omega }$ appears in the stochastic functional $L$. Then the phonon response
function can be written in the form 
\begin{equation}
G(k,\omega )=\langle \tilde{Q}_{{\bf -k},-\omega }Q_{{\bf k},\omega }\rangle
={\frac 1{{\cal Z}}}\int {\cal D}[S]{\cal D}[i\tilde{S}]{\cal D}[Q]{\cal D}[i%
\tilde{Q}]{\cal D}[e_S]{\cal D}[i\tilde{e}_S]{\cal D}[e_L]{\cal D}[i\tilde{e}%
_L]\tilde{Q}_{{\bf -k},-\omega }Q_{{\bf k},\omega }\exp (L)\ \ ,
\end{equation}
where ${\cal D}[\ ]$ denotes a suitable integration measure and ${\cal Z}$
is a normalization factor. The bilinear terms in the Lagrangian determine
the free response and correlation propagators. Usually, we include the terms
proportional to the coupling $w$ as well as the term $\gamma \tilde{e}_S%
\tilde{e}_L$ into the interaction part rather then into the free part of
Lagrangian. Then, the free propagators will be given by the expressions
presented in Table \ref{tab}, where $c=C_{11}^{1/2}$. Note that in order to
obtain the free response functions for the spin, spin-energy and
lattice-energy fields we have to multiply the response propagator by the
corresponding damping coefficient \cite{bausch}. Once the propagators are
defined in principle a perturbation expansion for any quantity can be
performed. Note also that because of the term $\gamma \tilde{e}_S\tilde{e}_L,
$ the spin-energy densities $\tilde{e}_S$, $e_S$ and the lattice-energy
densities $\tilde{e}_L$, $e_L$ are dynamically coupled, even if $w=0$ in
contrast to the static case where they are decoupled from each other in the
functional $H$ for $w=0$.

\subsection{Decoupling transformations}

We have found it more convenient first to decouple $\tilde{e}_S$, $e_S$ from 
$\tilde{e}_L$, $e_L$ by the Gaussian transformation: 
\[
e_{{\bf k},\omega }^L\rightarrow e_{{\bf k},\omega }^L+{\frac \gamma {%
C_S\Gamma _L}}D_{L0}(k,\omega )e_{{\bf k},\omega }^S+A(k,\omega )\tilde{e}_{%
{\bf k},\omega }^S\ \ \ ,
\]
\begin{equation}
\tilde{e}_{{\bf k},\omega }^L\rightarrow \tilde{e}_{{\bf k},\omega }^L+{%
\frac \gamma {C_L\Gamma _L}}D_{L0}(k,-\omega )\tilde{e}_{{\bf k},\omega }^S\
\ \ .  \label{a}
\end{equation}
Here $A(k,\omega )=-2\left[ {\frac \gamma {\Gamma _L}}D_{L0}(k,\omega )-{%
\frac \gamma {C_L\Gamma _L}}|D_{L0}(k,\omega )|^2\right] $ is a coefficient.
For the case of the total energy conserved, this transformation can be also
looked at as a substraction of the slow component associated with thermal
conduction mode from $e_L$, so the new variable $e_L$ contains only the fast
mode of equalization of temperatures: $e_L^{new}=C_L({\frac{e_L}{C_L}}-{%
\frac{e_S}{C_S}})$ for $\omega =0$. The Lagrangian expressed in new
variables takes the form

\begin{eqnarray}
L &=&\int_\omega \sum_{{\bf k}}\left\{ \Gamma \tilde{S}_{{\bf k},\omega }%
\tilde{S}_{-{\bf k},-\omega }+\theta k^2\tilde{Q}_{{\bf k},\omega }\tilde{Q}%
_{-{\bf k},-\omega }+\Gamma _{S0}(k,\omega )\tilde{e}_{{\bf k},\omega }^S%
\tilde{e}_{-{\bf k},-\omega }^S+\Gamma _L(k)\tilde{e}_{{\bf k},\omega }^L%
\tilde{e}_{-{\bf k},-\omega }^L\right.  \nonumber \\
&&-\tilde{Q}_{{\bf k},\omega }G_0^{-1}\left( k,\omega \right) Q_{{\bf k}%
,\omega }-\tilde{S}_{{\bf k},\omega }H_0^{-1}(k,\omega )S_{{\bf -k},-\omega
}-\tilde{e}_{{\bf k},\omega }^S\tilde{D}_{S0}^{-1}(k,\omega )e_{-{\bf k}%
,-\omega }^S-\tilde{e}_{{\bf k},\omega }^LD_{L0}^{-1}(k,\omega )e_{-{\bf k}%
,-\omega }^L  \nonumber \\
&&-u\Gamma \tilde{S}_{{\bf k},\omega }^2S_{-{\bf k},-\omega }^2-2f\Gamma 
\widetilde{S^2}_{{\bf k},\omega }e_{{\bf -k},-\omega }^S-fN(k,\omega )\Gamma
_S(k)\tilde{e}_{{\bf k},\omega }^SS_{-{\bf k},-\omega }^2+f\gamma \tilde{e}_{%
{\bf k},\omega }^LS_{-{\bf k},-\omega }^2 \\
&&-iwkJ(k,\omega )\tilde{Q}_{{\bf k},\omega }e_{{\bf -k},-\omega
}^S-iawkA_{k,-\omega }\tilde{Q}_{{\bf k,}\omega }\tilde{e}_{{\bf -k},-\omega
}^S-ikwK(k,\omega )\Gamma _S(k)\tilde{e}_{{\bf k},\omega }^SQ_{-{\bf k}%
,-\omega }  \nonumber \\
&&\left. -iw(a\Gamma _L(k)-\gamma )k\tilde{e}_{{\bf k},\omega }^LQ_{-{\bf k}%
,-\omega }-iawk\tilde{Q}_{{\bf k,}\omega }e_{{\bf -k},-\omega }^L-2igk\Gamma 
\widetilde{S^2}_{{\bf k},\omega }Q_{-{\bf k},-\omega }-igk\tilde{Q}_{{\bf k,}%
\omega }S_{-{\bf k},-\omega }^2\right\} \ ,  \nonumber
\end{eqnarray}
where $\widetilde{S^2}_{{\bf k},\omega }=(\tilde{S}S)_{{\bf k},\omega }$, $%
S_{{\bf k},\omega }^2=(S^2)_{{\bf k},\omega }$ and

\[
\Gamma _{S0}\left( k,\omega \right) =\Gamma _S(k)-\Gamma _L\left( k\right) 
\frac{\gamma ^2\mid D_{L0}(k,\omega )\mid ^2}{C_L^2}\ \ , 
\]
$\ \ $%
\[
N(k,\omega )=1-\frac{\gamma ^2D_{L0}(k,\omega )}{C_L\Gamma _S}\ \
,J(k,\omega )=1+{\frac{a\gamma D_{L0}(k,\omega )}{C_S}}\ \ , 
\]
\[
K(k,\omega )=N(k,\omega )-\frac{a\gamma }{\Gamma _S}(1-\frac{D_{L0}(k,\omega
)\Gamma _L}{C_L})\ . 
\]
As a result of the transformation (\ref{a}) the spin-energy response
propagator is dressed to the form 
\begin{equation}
\tilde{D}_{S0}^{-1}(k,\omega )=D_{S0}^{-1}(k,\omega )-{\frac{\gamma
^2D_{L0}(k,\omega )}{\Gamma _SC_SC_L}}=C_S^{-1}\Gamma _S\left( 1-i\tilde{%
\omega}-\frac{\gamma ^2}{\Gamma _S\Gamma _L\left( 1-i\bar{\omega}\right) }%
\right)  \label{dressed}
\end{equation}
where $\tilde{\omega}=\omega /\omega _S$ and $\bar{\omega}=\omega /\omega _L$
is the frequency reduced with respect to the bare spin-energy $\omega
_S=\Gamma _S/C_S$ and lattice-energy $\omega _L=\Gamma _L/C_L$ relaxation
rate, respectively. It is easy to see that for very slow lattice-energy
relaxation $\bar{\omega}\gg 1$, which takes place for $\gamma _L=\gamma
\rightarrow 0$ or $C_L\rightarrow \infty $, the dressed spin-energy
propagator reduces to the bare one and the model with two energies reduces
to the model without the lattice-energy field, in which only the equation

\[
\dot{e}_S=-\Gamma _S(k)\frac{\delta H}{\delta e_S}+\varphi 
\]
matters in our considerations. Depending on the ratio of the bare relaxation
rate of the energy diffusion mode $\frac{\lambda _Sk^2}{C_S}$ to the bare
spin-lattice relaxation frequency $\gamma /C_S,$ the spin system can be then
regarded either as thermally isolated from the lattice ($\frac{\lambda _Sk^2}%
\gamma \gg 1$) or as freely relaxing to an infinite heat bath at each
lattice site, or to a bath with an infinite thermal conductivity ($\frac{%
\lambda _Sk^2}\gamma \ll 1$). The same happens for the nonconserved energy
case. If $\gamma _L\rightarrow \infty $ the lattice energy always takes its
equilibrium value. Then $\bar{\omega}\ll 1$, $\frac{\gamma ^2}{\Gamma
_L\Gamma _S}\rightarrow 0$ and again $\tilde{D}_{S0}\rightarrow D_{S0}.$

The one energy field model we recover also for $\gamma \rightarrow \infty $
or $C_L\rightarrow 0$ in the conserved total energy case. Then $\bar{\omega}%
\ll 1$ and expanding Eq.(\ref{dressed}) we get $\tilde{D}_{S0}(k,\omega
)=C_S\Gamma _S^{-1}m^{-1}\left( 1-i\frac{\omega C_{+}}{\lambda _{+}k^2}%
\right) ^{-1}$ where $m(k)\equiv 1-{\frac{\gamma ^2}{\Gamma _S(k)\Gamma _L(k)%
}\simeq }\frac{\lambda _{+}k^2}\gamma $ with $\lambda _{+}=\lambda
_S+\lambda _L$ and $C_{+}=C_S+C_L$. being the thermal conductivity and
specific-heat of the total $S+L$ system. At the same time the dressed
damping coefficient $\Gamma _{S0}\left( k,\omega \right) \rightarrow \Gamma
_S(k)m$ so the dressed spin-energy response function $\Gamma _{S0}\left(
k,\omega \right) \tilde{D}_{S0}(k,\omega )\rightarrow C_S\left( 1-i\frac{%
\omega C_{+}}{\lambda _{+}k^2}\right) ^{-1}$. It can be said that on very
short time scales, of an order of $C_L/\gamma ,$ the lattice comes to
equilibrium with the spin system and then for much longer times, of an order
of the inverse of ultrasonic frequency, only the diffusion of the total
energy $e_S+e_L$ matters and instead of Eqs.(\ref{2c},\ref{2d}) only the
equation

\[
\dot{e}=-\lambda _{+}k^2\frac{\delta H}{\delta e}+\phi 
\]
may be used. This limiting case for dynamics was considered by Drossel and
Schwabl \cite{dros93}. The reason for introducing the second energy field is
now becoming clear. We expect that the model with only one energy density
correctly describes the dynamics only in the two limiting cases: for pure
diffusion of the energy field and for pure relaxation to an infinite heat
bath. As we want to be able also to describe the dynamics which is
intermediate between these two cases we have included $e_L$ as a separate
variable.

Next, we decouple the sound mode from the spin and energy fluctuations by
the transformation 
\begin{eqnarray*}
Q_{{\bf k},\omega }\rightarrow Q_{{\bf k},\omega } &-&gkG_0(k,\omega )S_{%
{\bf k},\omega }^2-wkJ(k,\omega )G_0(k,\omega )e_{{\bf k},\omega
}^S-awkG_0(k,\omega )e_{{\bf k},\omega }^L \\
&&+B(k,\omega )\widetilde{S^2}_{{\bf k},\omega }+E(k,\omega )\tilde{e}_{{\bf %
k},\omega }^S+F(k,\omega )\tilde{e}_{{\bf k},\omega }^L\ \ \ ,
\end{eqnarray*}
\begin{equation}
\tilde{Q}_{{\bf k},\omega }\rightarrow \tilde{Q}_{{\bf k},\omega
}-2gkG_0(k,-\omega )\Gamma \widetilde{S^2}_{{\bf k},\omega }-wkK(k,-\omega
)\Gamma _SG_0(k,-\omega )\tilde{e}_{{\bf k},\omega }^S-wk(a\Gamma _L-\gamma
)G_0(k,-\omega )\tilde{e}_{{\bf k},\omega }^L\ \ \ .  \label{b}
\end{equation}
Here the abbreviations 
\[
B(k,\omega )=-4\Gamma gk|G_0|^2\theta k^2\ \ ,
\]
\[
E(k,\omega )=-2wkK(k,-\omega )\Gamma _S|G_0|^2\theta k^2-awkA(k,\omega
)G_0(k,\omega )\ \ ,
\]
\[
F(k,\omega )=-2wk(a\Gamma _L-\gamma )|G_0|^2\theta k^2\ \ \ ,
\]
have been introduced. Finally, we apply the transformations: 
\[
e_{{\bf k},\omega }^L\rightarrow e_{{\bf k},\omega }^L+R(k,\omega )S_{{\bf k}%
,\omega }^2+T(k,\omega )e_{{\bf k},\omega }^S+U(k,\omega )\widetilde{S^2}_{%
{\bf k},\omega }+V(k,\omega )\tilde{e}_{{\bf k},\omega }^S\ \ \ ,
\]
\begin{equation}
\tilde{e}_{{\bf k},\omega }^L\rightarrow \tilde{e}_{{\bf k},\omega }^L+%
\tilde{D}_{L0}(k,-\omega )\left[ 2awgk^2G_0(k,-\omega )\Gamma \widetilde{S^2}%
_{{\bf k},\omega }+aw^2k^2K(k,-\omega )G_0(k,-\omega )\Gamma _S\tilde{e}_{%
{\bf k},\omega }^S\right] \ \ \ ,  \label{c}
\end{equation}
and 
\[
e_{{\bf k},\omega }^S\rightarrow e_{{\bf k},\omega }^S+\tilde{\tilde{D}}%
_{S0}(k,\omega )\Gamma _S\left[ wgk^2K(k,\omega )G_0(k,\omega )-fN(k,\omega
)\right] S_{{\bf k},\omega }^2+W(k,\omega )\widetilde{S^2}_{{\bf k},\omega
}\ \ \ ,
\]
\begin{equation}
\tilde{e}_{{\bf k},\omega }^S\rightarrow \tilde{e}_{{\bf k},\omega }^S-2%
\tilde{\tilde{D}}_{S0}(k,-\omega )\left[ f-wgk^2J(k,-\omega )G_0(k,-\omega )\right]
\Gamma \widetilde{S^2}_{{\bf k},\omega }\ \ \ ,  \label{d}
\end{equation}
decoupling $e_S$ and $e_L$ from the spin fluctuations, where 
\[
\tilde{D}_{L0}^{-1}(k,\omega )=D_{L0}^{-1}(k,\omega )-(a\Gamma _L-{\gamma }%
)aw^2k^2G_0(k,\omega )\ \ \ ,
\]
\[
\tilde{\tilde{D}}_{S0}^{-1}(k,\omega )=\tilde{D}_{S0}^{-1}(k,\omega
)-w^2g^2k^2J(k,\omega )K(k,\omega )G_0(k,\omega )\ \ \ ,
\]
are ``dressed'' by transformations (\ref{a})\ and\ (\ref{b}) energy-density
propagators. Here the coefficients are given by 
\[
R(k,\omega )=\left[ {\gamma }f+(a\Gamma _L-{\gamma })wgk^2G_0(k,\omega
)\right] \tilde{D}_{L0}(k,\omega )\ \ \ ,
\]
\[
T(k,\omega )=w^2k^2J(k,\omega )(a\Gamma _L-{\gamma })G_0(k,\omega )\tilde{D}%
_{L0}(k,\omega )\ \ \ ,
\]
\[
U(k,\omega )=4\Gamma wgk^2(a\Gamma _L-{\gamma })|G_0|^2\theta k^2\tilde{D}%
_{L0}(k,\omega )+4agwk^2\Gamma \Gamma _LG_0(k,-\omega )|\tilde{D}_{L0}|^2\ \
\ ,
\]
\[
V(k,\omega )=wk\left\{ 2awkK(k,-\omega )[1+(a\Gamma _L-{\gamma }%
)w^2k^2|G_0|^2\theta k^2]G_0(k,-\omega )|\tilde{D}_{L0}|^2\Gamma _S\Gamma
_L-(a\Gamma _L-\gamma )E(k,-\omega )\tilde{D}_{L0}(k,\omega )\right\} \ ,
\]
\begin{eqnarray*}
W(k,\omega ) &=&2wgk^2\left[ aG_0(k,-\omega )A(k,\omega )+2K(k,\omega
)\Gamma _S|G_0|^2\theta k^2\right] \Gamma \tilde{\tilde{D}}_{S0}(k,\omega ) \\
&&-4\left[ f-wgk^2J(k,-\omega )G_0(k,-\omega )\right] \left( 1-{\frac{\gamma
^2\Gamma _L\Gamma |D_{L0}|^2}{\Gamma _SC_L^2}}\right) {\Gamma \Gamma }_S|%
\tilde{\tilde{D}}_{S0}|^2\ \ \ .
\end{eqnarray*}

\section{ACOUSTIC SELF-ENERGY}

Having applied the transformations (\ref{a}, \ref{b}-\ref{d}) the acoustic
phonon response function, to the leading order in coupling constants $g$ and 
$w$, can be written as 
\begin{equation}
\langle \tilde{Q}_{{\bf -k},-\omega }Q_{{\bf k},\omega }\rangle
=G_0(k,\omega )+G_0^2(k,\omega )\Sigma ^{\text{red}}(k,\omega )\ \ ,
\label{sigmar}
\end{equation}
where 
\begin{equation}
\Sigma ^{\text{red}}(k,\omega )=k^2\left\{ X(k,\omega )\tilde{D}%
_{L0}(k,\omega )+Y(k,\omega )\tilde{\tilde{D}}_{S0}(k,\omega )+2g_1(k,\omega
)g_2(k,\omega )\langle \Gamma \widetilde{S^2}_{{\bf -k},-\omega }S_{{\bf k}%
,\omega }^2\rangle ^{L_{\text{eff}}}\right\} ,  \label{efflag}
\end{equation}
and 
\[
X(k,\omega )=a(a\Gamma _L-{\gamma })w^2\ \ \ \ ,\ \ \ Y(k,\omega )=w^2\Gamma
_SK(k,\omega )J(k,\omega ),
\]
\[
g_1(k,\omega )=g-w\hat{f}_1\Gamma _SK(k,\omega )\tilde{\tilde{D}}_{S0}(k,\omega )+{aw%
\tilde{f}\Gamma }_L{\tilde{D}_{L0}(k,\omega )}\ \ \ ,
\]
\[
g_2(k,\omega )=g-w\hat{f}_2J(k,\omega )\Gamma _S\tilde{\tilde{D}}_{S0}(k,\omega )\ \
\ ,
\]
with $\tilde{f}={\gamma }f+(a\Gamma _L-{\gamma })wgk^2G_0(k,\omega )$, $\hat{%
f}_1=fN(k,\omega )-wgk^2K(k,\omega )G_0(k,\omega )$ and $\hat{f}%
_2=f-wgk^2J(k,\omega )G_0(k,\omega )$. The crucial point in Eq.(\ref{efflag}%
) is that the four-spin response function is calculated with the Lagrangian
which depends solely on the spin variables. As the result of coupling of the
spin variable to the energy densities and phonons, the effective spin
Lagrangian $L_{\text{eff}}$ obtained by transformations (\ref{a}, \ref{b}-%
\ref{d}) contains phonon- as well as energy-density-mediated four-spin
non-local interactions %the most important of which is 
$u(k,\omega )\widetilde{S^2}_{{\bf -k},-\omega }S_{{\bf k},\omega }^2$ with 
%the renormalised coupling  
\begin{equation}
u(k,\omega )=u-2g^2k^2G_0(k,\omega )-2awgk^2\tilde{f}\Gamma _LG_0(k,\omega )%
\tilde{D}_{L0}(k,\omega )-2\hat{f}_1\hat{f}_2\Gamma _S\tilde{\tilde{D}}%
_{S0}(k,\omega )\ \ .  \label{u}
\end{equation}
A general expression for the interacting phonon response function is given by

\begin{equation}
G^{-1}(k,\omega )=G_0^{-1}(k,\omega )-\Sigma (k,\omega ),  \label{siggen}
\end{equation}
where the phonon self-energy $\Sigma (k,\omega )$ is the irreducible, with
respect to phonon lines, part of $\Sigma ^{\text{red}}(k,\omega )$. From
Eqs.\ (\ref{sigmar}) and (\ref{siggen}) the irreducible and reducible parts are
related through 
\begin{equation}
\Sigma (k,\omega )={\frac{\Sigma ^{\text{red}}(k,\omega )}{1+G_0(k,\omega
)\Sigma ^{\text{red}}(k,\omega )}}.  \label{sigred}
\end{equation}
Next we may eliminate the dangerous resonances \cite{deng,pawepj} by
replacing $G_0(k,\omega )$ with $G_0(k,0)$ in $\Sigma (k,\omega )$ i.e. we
set the strongly irrelevant parameters - the coefficients at $\omega k^2$
and $\omega ^2$ - equal to zero. In this paper we will not discuss the role
of the denominator in Eq.\ (\ref{sigred}) assuming the elastic couplings $g$
and $w$ to be very small. This assumption gives $\Sigma (k,\omega )=\Sigma ^{%
\text{red}}(k,\omega )$. It is consistent with neglecting the higher order
(in $g$ and $w$) terms in $\Sigma ^{\text{red}}(k,\omega )$. Then also $%
\tilde{D}_{L0}(k,\omega )=D_{L0}(k,\omega )$ and $\tilde{\tilde{D}}_{S0}(k,\omega )=%
\tilde{D}_{S0}(k,\omega )$ together with small $g$ and $w$ let us also
neglect the macroscopic instability \cite{paw89,berg}, which is believed to
take place in compressible spin systems with positive specific-heat exponent
near the transition temperature. The weak first-order transition in such
systems is a result of a non-analytical character of the coupling constant $%
u(k,\omega )$ at $k=0$. For small $g$ and $w$ the first-order regime can be
probably observed only extremely close to $T_C$. We could in principle put
other irrelevant parameter in Eq.\ (\ref{sigred}) equal to zero as the
coefficient at $\omega $ in the energy-density propagators for non-conserved
systems but then we would obtain only the asymptotic behavior of the sound
attenuation coefficient. As we are also interested in the nonasymptotic
behavior associated with the coupling to the energy-density we will proceed
in another way, preserving the nonasymptotic effects. First, note, that
because for small $g$ and $w$ 
\[
u(k,\omega )=u-2f^2N(k,\omega )\Gamma _S\tilde{D}_{S0}(k,\omega ),
\]
thus $\langle \Gamma \widetilde{S^2}_{{\bf -k},-\omega }S_{{\bf k},\omega
}^2\rangle ^{L_{\text{eff}}}$ is reducible with respect to (new) spin-energy
propagators and so is $\Sigma $ . We assume that the coupling constant $f$
need not to be very small but the positivity of the four-spin coupling must
be guaranteed \cite{pawepj}. % can take arbitrary values 
The perturbation expansion with respect to $u(k,\omega )$ let us express $%
\langle \Gamma \widetilde{S^2}_{{\bf -k},-\omega }S_{{\bf k},\omega
}^2\rangle ^{L_{\text{eff}}}$, the reducible with respect to $\tilde{D}_S$
four-spin response function, in terms of the irreducible one 
\begin{equation}
\langle \Gamma \widetilde{S^2}_{{\bf -k},-\omega }S_{{\bf k},\omega
}^2\rangle ^{L_{\text{eff}}}=\frac{\langle \Gamma \widetilde{S^2}_{{\bf -k}%
,-\omega }S_{{\bf k},\omega }^2\rangle ^{\text{irr}}}{1-2f^2\Gamma
_SN(k,\omega )\tilde{D}_{S0}(k,\omega )\langle \Gamma \widetilde{S^2}_{{\bf %
-k},-\omega }S_{{\bf k},\omega }^2\rangle ^{\text{irr}}}.  \label{irr}
\end{equation}
Eq.\ (\ref{irr}) is equivalent to 
\begin{equation}
\langle \Gamma \widetilde{S^2}_{{\bf -k},-\omega }S_{{\bf k},\omega
}^2\rangle ^{\text{irr}}=\frac{\langle \Gamma \widetilde{S^2}_{{\bf -k}%
,-\omega }S_{{\bf k},\omega }^2\rangle ^{L_{\text{eff}}}}{1+2f^2\Gamma
_SN(k,\omega )\tilde{D}_{S0}(k,\omega )\langle \Gamma \widetilde{S^2}_{{\bf %
-k},-\omega }S_{{\bf k},\omega }^2\rangle ^{L_{\text{eff}}}}.  \label{irr2}
\end{equation}
It is easy to note that the coefficient preceding $\langle \Gamma \widetilde{%
S^2}_{{\bf -k},-\omega }S_{{\bf k},\omega }^2\rangle ^{L_{\text{eff}}}$ in
the denominator of Eq.\ (\ref{irr2}) can be written as $2f^2\hat{D}%
_S(k,\omega )$ with $\hat{D}_S(k,\omega )=C_S\left( 1-i\tilde{\omega}%
b(q,\omega )\right) ^{-1}$ where $b(q,\omega )={\frac{1-i\bar{\omega}}{m-i%
\bar{\omega}}}$. In the $\omega =0$ limit it is equal to $v=2f^2C_S$. As $%
\langle \Gamma \widetilde{S^2}_{{\bf -k},-\omega }S_{{\bf k},\omega
}^2\rangle ^{\text{irr}}$ is a vertex function, containing no resonances, we
may again replace strongly irrelevant parameters in Eq.\ (\ref{irr2}) like
the coefficient at $\omega $ in $\tilde{D}_{S0}$, in the case of
nonconserved systems, by zero. We then obtain 
\begin{equation}
\langle \Gamma \widetilde{S^2}_{{\bf -k},-\omega }S_{{\bf k},\omega
}^2\rangle ^{\text{irr}}=\frac{\langle \Gamma \widetilde{S^2}_{{\bf -k}%
,-\omega }S_{{\bf k},\omega }^2\rangle ^{L_{\text{A}}}}{1+v\langle \Gamma 
\widetilde{S^2}_{{\bf -k},-\omega }S_{{\bf k},\omega }^2\rangle ^{L_{\text{A}%
}}}\hspace{1cm},  \label{irr3}
\end{equation}
where $L_{\text{A}}$ is the action of the model A of Halperin, Hohenberg and
Ma \cite{hhm74}, with $u_{\text{A}}=u-v$. In the case of a system with the
conserved total energy we cannot proceed in this way as $\tilde{D}_{S0}$
contains nonstatic relevant terms and $\langle \Gamma \widetilde{S^2}_{{\bf %
-k},-\omega }S_{{\bf k},\omega }^2\rangle ^{\text{irr}}$ should be
calculated using energy generated dynamic interactions in $L_{\text{eff}}$ 
\cite{hhm76}. The asymptotic behavior of the ultrasonic attenuation
coefficient in a model with conserved energy field was investigated by
Drossel and Schwabl \cite{dros93}. They showed that the difference of the
asymptotic scaling functions (compared to the model A) originates mainly
from different exponent $z$ for model C. In the present paper we will
restrict our discussion to the nonconserved systems and a detailed analysis
of the conserved systems will be presented elsewhere.

Now we return to Eq.\ (\ref{irr3}) which inserted into Eq.\ (\ref{irr})
yields 
\begin{equation}
\langle \Gamma \widetilde{S^2}_{{\bf -k},-\omega }S_{{\bf k},\omega
}^2\rangle ^{L_{\text{eff}}}=\frac{\langle \Gamma \widetilde{S^2}_{{\bf -k}%
,-\omega }S_{{\bf k},\omega }^2\rangle ^{L_{\text{A}}}}{1+2f^2(C_S-\hat{D}%
_S)\langle \Gamma \widetilde{S^2}_{{\bf -k},-\omega }S_{{\bf k},\omega
}^2\rangle ^{L_{\text{A}}}}\ \ \ .  \label{irr4}
\end{equation}
Next from Eqs. (\ref{efflag}),(\ref{sigred}) and (\ref{irr4}) we obtain 
\begin{equation}
\Sigma (k,\omega )=k^2\left\{ X(k,\omega )D_{L0}(k,\omega )+Y(k,\omega )%
\tilde{D}_{S0}(k,\omega )+\frac{2g_1(k,\omega )g_2(k,\omega )\langle \Gamma 
\widetilde{S^2}_{{\bf -k},-\omega }S_{{\bf k},\omega }^2\rangle ^{L_{\text{A}%
}}}{1+2f^2(C_S-\hat{D}_S)\langle \Gamma \widetilde{S^2}_{{\bf -k},-\omega
}S_{{\bf k},\omega }^2\rangle ^{L_{\text{A}}}}\right\} .  \label{sigma}
\end{equation}
It is not difficult to note that 
\[
g_1(k,\omega )=g_2(k,\omega )=\hat{g}(k,\omega )=g-wfP(k,\omega )\hat{D}%
_S(k,\omega ), 
\]
with

\[
P(k,\omega )={\frac{m-i\bar{\omega}(1-\frac{a\gamma }{\Gamma _S})}{m-i\bar{
\omega}}}. 
\]
Now let us split $AD_{L0}+B\tilde{D}_{S0}$ into the following terms 
\[
{\frac{a^2w^2k^2mC_L}{m-i\bar{\omega}}}+w^2k^2P^2\hat{D}_S(k,\omega ) 
\]
and reduce the second term to the common denominator with the third term in
Eq.\ (\ref{sigma}). This results in the following expression for $\Sigma $ 
\begin{equation}
{\frac{\Sigma (k,\omega )}{k^2}}={\frac{a^2w^2C_Lm}{m-i\bar{\omega}}}+\frac{%
[2w^2f^2P^2C_S\hat{D}_S(k,\omega )-4wfgP\hat{D}_S(k,\omega )+2g^2]\langle
\Gamma \widetilde{S^2}_{{\bf -k},-\omega }S_{{\bf k},\omega }^2\rangle ^{L_{%
\text{A}}}+w^2P^2}{1+2f^2(C_S-\hat{D}_S)\langle \Gamma \widetilde{S^2}_{{\bf %
-k},-\omega }S_{{\bf k},\omega }^2\rangle ^{L_{\text{A}}}}.  \label{sigma2}
\end{equation}
Remembering that $C_S-\hat{D}_S(k,\omega )=-i\tilde{\omega}b(k,\omega )\hat{D%
}_S(k,\omega )$ and ignoring the first noncritical term in Eq.\ (\ref{sigma2}%
) the acoustic self-energy can be written as 
\begin{equation}
{\frac{\Sigma (k,\omega )}{k^2}}=\frac{2[g^2(k,\omega )-i\tilde{\omega}%
b(k,\omega )g^2]\langle \Gamma \widetilde{S^2}_{{\bf -k},-\omega }S_{{\bf k}%
,\omega }^2\rangle ^{L_{\text{A}}}+w^2P^2C_S}{1-i\tilde{\omega}b(k,\omega
)(1+v\langle \Gamma \widetilde{S^2}_{{\bf -k},-\omega }S_{{\bf k},\omega
}^2\rangle ^{L_{\text{A}}})}\ \ .  \label{sigma3}
\end{equation}
with $g(k,\omega )=g-wfC_SP(k,\omega )$.

\section{DISCUSSION}

\subsection{General expressions}

The critical contribution to the coefficient of attenuation is determined by
the imaginary part of $\Sigma $ and Eq.\ (\ref{sigma3}) implies 
\begin{equation}
\alpha (\omega ,t)=({\frac \omega {2c^3})}{\frac{W_1|\Psi |^2+W_2{\rm Im}%
\Psi +W_3{\rm Re}\Psi }{|1-i\tilde{\omega}b(1+v\Psi )|^2}}  \label{alfa}
\end{equation}
where $\Psi =\langle \Gamma \widetilde{S^2}_{{\bf -k},-\omega }S_{{\bf k}%
,\omega }^2\rangle ^{L_{\text{A}}}$ and $W_1=2v\tilde{\omega}{\rm Re}[\bar{b}%
g^2(k,\omega )]$, 
\[
W_2=2{\rm Re}g^2(k,\omega )+2\tilde{\omega}^2|b|^2g^2+\tilde{\omega}{\rm Im}%
[v\bar{b}w^2C_SP^2+2bg^2-2\bar{b}g^2(k,\omega )], 
\]
\[
W_3=2\tilde{\omega}{\rm Re}[\bar{b}g^2(k,\omega )-bg^2+\bar{b}%
w^2f^2C_S^2P^2]+2{\rm Im}g^2(k,\omega ). 
\]
In the limit $\Gamma _L\rightarrow \infty $ these coefficients take the
corresponding values from the model without lattice-energy field \cite
{pawepj} i.e. $W_1=2v\tilde{\omega}\tilde{g}^2$, $W_2=2(\tilde{g}^2+\tilde{%
\omega}^2g^2)$ and $W_3=-4wfC_S\tilde{\omega}\tilde{g}$ with $\tilde{g}%
=g-wfC_S$. The four-spin response function usually is evaluated in the limit 
$k=0$ since in the ultrasonic experiments the wavelength is much longer than
the correlation length ($k\xi \ll 1$) while the ultrasonic frequency can be
comparable to the characteristic frequency of the spin fluctuations. Then
the singular part of the four-spin response function $\Psi $ obeys the
scaling relation 
\[
\Psi =t^{-\alpha }\Phi (y)\hspace{1cm}, 
\]
with $y=\omega t^{-z\nu }/\Gamma $ as the reduced frequency and $t$
proportional to the reduced temperature. Here $\alpha ,\nu $ and $z$ are
usual critical exponents. The scaling function $\Phi $ is known to the
leading order in $\epsilon =4-d$ \cite{iros,paw89,deng} : 
\[
\Phi (y)=\Theta ^{-\alpha /\nu }\left\{ \frac \nu \alpha +\frac iy\left[
i\left( 1-iy/2\right) \arctan (y/2)-\frac 12\ln \left( 1+(y/2)^2\right)
\right] \right\} K_4, 
\]
where $\Theta =[1+(y/2)^2]^{-{\frac 14}}$ and $K_4=(8\pi ^2)^{-1}$. With the
aid of Eq.\ (\ref{alfa}) the attenuation coefficient can be written as 
\begin{equation}
\alpha (\omega ,t)={\frac{\omega ^2}{2c^3}}\left[ \frac{C_S}{\Gamma _S}%
\tilde{W}_1t^{-2\alpha }F_1(y;\omega ,t)+\frac 1\Gamma \tilde{W}%
_2t^{-(\alpha +z\nu )}F_2(y;\omega ,t)+\tilde{W}_3t^{-\alpha }F_3(y;\omega
,t)\right] ,  \label{alfa2}
\end{equation}
where $\tilde{W}_1=W_1K_4/\tilde{\omega}$, $\tilde{W}_2=W_2K_4$, $\tilde{W}%
_3=W_3K_4/\omega $ and $F_1=|\Phi /M|^2$, $F_2={\rm Im}(\Phi )y^{-1}|M|^{-2}$%
, $F_3={\rm Re}(\Phi )|M|^{-2}$ with the denominator of Eq.(\ref{sigma3})
denoted by 
\[
M(t,\omega )=1-i\tilde{\omega}b(1+vt^{-\alpha }\Phi ). 
\]

\subsection{Regime I}

Eq.\ (\ref{alfa2}) shows many different regimes. First let us consider the
regime I: $m\gg \bar{\omega}$ ($\omega \ll \frac{(\gamma _{L-}\gamma
)+\lambda _{+}k^2}{C_L}$) for which $P(\omega )\rightarrow 1$. It seems that
the regime I may be adequate for some metals, where $e_L$ should be
understood as the spin energy of the conduction electrons and also infinite
specific-heat associated with phonons can be assumed.

\subsubsection{Low-frequency region}

In the low-frequency region, $\tilde{\omega}|b|\simeq \frac{\tilde{\omega}}m%
\ll 1$, the denominator in Eq.\ (\ref{alfa}) does not play any role and the
second term in the numerator with strong singularity dominates 
\begin{equation}
\alpha (\omega ,t)\simeq \frac{\tilde{g}^2}{c^3\Gamma }\omega ^2t^{-(z\nu
+\alpha )}{\frac{{\rm Im}\Phi (y)}y}.  \label{asym}
\end{equation}
This type of singularity in the ultrasonic attenuation, which we shall also
call Murata-Iro-Schwabl behavior, has been so far obtained by neglecting the
energy density fields \cite{mur,iros} and it is believed to take place in
magnetic metals \cite{lut,kaw76}. Note, however that the coefficient $W_2$
in this limit is equal $2(g-wfC_S)^2$ to be compared with $2g^2$ in the
model without coupling to the energy fields \cite{iros}.

When $\frac{\tilde{\omega}}m\ll 1$ i.e. for $M(t,\omega )\simeq 1$ while $t$
is not extremely small there may be a competition between the first and the
second term in Eq.\ (\ref{alfa2}). For $t>t_{\text{cross}}$ with $t_{\text{%
cross}}=\left( {\frac{\alpha ^2m\gamma _S}{4K_4\nu ^2vC_S\Gamma }}\right) ^{%
\frac 1{z\nu -\alpha }}$ the weak-singularity term dominates and 
\begin{equation}
\alpha (\omega ,t)\simeq \frac{\tilde{g}^2vC_S}{c^3 m\gamma _S}\omega
^2t^{-2\alpha }|\Phi (y)|^2.  \label{kawbeh}
\end{equation}
Such behavior was first obtained by Kawasaki \cite{kaw69}. He postulated
phonon-spin-energy coupling in order to explain extremely small sound
attenuation exponents observed in magnetic insulators. Note however the
difference between the corresponding scaling functions. What also arises
from our analysis, is that such behavior is generated even if there is no
direct coupling of the sound mode to the energy-density fields, and that it
cannot be truly asymptotic as both terms (\ref{asym}) and (\ref{kawbeh}) are
proportional (in the $\omega \rightarrow 0$ limit) to the square of the same
coupling constant $\tilde{g}$ so even if $g=0$ the strong singularity
described by Eq.\ (\ref{asym}) is still present and become dominant for
sufficiently small $t$. The dominance of a given type of behavior does not
depend on the relative strength of the coupling constants $g$ and $w$ but
rather on the ratio of the bare relaxation times ${\frac{\gamma _Sm}{%
C_S\Gamma }}$ and also on the ratio of $v$ to its fixed point value $%
v^{*}\simeq (\alpha /\nu )K_4^{-1}$. Slow spin-lattice relaxation (and/or
fast decay of spin fluctuations) together with strong coupling $v$ favors
the weak-singularity type behavior.

It should be noted that the amplitudes of the Murata-Iro-Schwabl as well as
the Kawasaki term are proportional to the same coupling constant $\tilde{g}%
^2 $ only in the limit $\tilde{\omega}=\bar{\omega}=0$. For finite $\tilde{%
\omega}$ or $\bar{\omega}$ both amplitudes begin to differ from each other
and become frequency dependent as can be seen from the expressions for the
coefficients $\tilde{W}_1$and $\tilde{W}_2$. For example $\tilde{W}_1$ is
not longer equal to $\tilde{W}_2-2\tilde{\omega}^2|b|^2g^2$. To be exact
this frequency-dependence (except $2\tilde{\omega}^2|b|^2g^2$ term in $%
\tilde{W}_2$) of the coefficients $\tilde{W}_i$ results from the inclusion
of the lattice-energy field to the model in which spins, phonons and
spin-energy modes interact. The nonuniversal frequency-dependent
coefficients may be useful in making comparisons with experimental results.

\subsubsection{High-frequency region}

As the sound frequency increases, $\tilde{\omega}/m$ approaches unity and
the denominator in Eq.\ (\ref{alfa}) as well as the second term in the
expression for $W_2$ become important. With a further increase of the sound
frequency, two scenarios, can be realized. In the first one, $\tilde{W}_2\ll 
\tilde{W}_1$ i.e. the sound frequency is not high enough and/or $g$ is much
smaller then $\tilde{g}$. Then, the situation resembles the high frequency
regime for Kawasaki classical relaxation function \cite{kaw71}. The sound
attenuation coefficient saturates and $\alpha (\omega ,t)\propto {\rm const}$%
. The second scenario happens when $\tilde{W}_2$ in which the term $2g^2%
\tilde{\omega}^2|b|^2$ dominates, is much greater then $\tilde{W}_1$ and
then:

(i) for $vt^{-\alpha }\Phi \gg 1$ i.e. with noticeable singularity in the
specific-heat a new type of behavior can be observed 
\begin{equation}
\alpha (\omega ,t)\propto g^2\omega ^2t^{-(z\nu -\alpha )}\frac{{\rm Im}\Phi
(y{)^{-1}}}y,  \label{mod}
\end{equation}
which can be obtained from the one described be Eq.\ (\ref{asym}) by a
simple replacement $\alpha \rightarrow -\alpha $, $\Phi \rightarrow \Phi
^{-1}$ and $\tilde{g}\rightarrow g$ giving a bit weaker than
Murata-Iro-Schwabl singularity but much stronger than the anomaly in the
energy-relaxation-dominated region described by Eq.(\ref{kawbeh}). The
difference in exponents $z\nu +\alpha $ and $z\nu -\alpha $ is very small
and it may be difficult to distinguish both regimes in experiment. This new
type of behavior of ultrasonic attenuation in magnets is very interesting.
In this regime the frequency of the sound mode is much higher than the
(modified by the lattice) spin-energy relaxation frequency so contrary to
the low-frequency region described by Eqs.\ (\ref{asym}) and (\ref{kawbeh})
now the ``lattice'' and the spin systems no longer can be treated as
remaining at a local equilibrium. The local temperature of the spin system
is not able to follow alternate hot and cold temperature variations produced
by the ultrasonic wave. One can say that we have here some kind of
``adiabatic'' sound propagation although the ``lattice'' (whatever it is in
regime I), in virtue of the assumption $\bar{\omega}\ll m<1$ defining the
region I, can be treated as at equilibrium with the ultrasonic wave.

The behavior described by Eq.\ (\ref{asym}) can be regarded as dominant for
the whole asymptotic region $t,\omega \rightarrow 0$ except only for a very
narrow area $vt^{-\alpha }\tilde{\omega}/m\gg 1$ where the ``adiabatic''
behavior (\ref{mod}) is dominant. The latter area, however, can be 
experimentally inaccessible as for sufficiently low sound-frequency the
reduced temperature needed to make $vt^{-\alpha }\tilde{\omega}/m$ big
enough may be too small, as a consequence of smallness of the exponent $%
\alpha $, to be approached in real experiment. For this reason we shall
sometimes call the Murata-Iro-Schwabl behavior also the asymptotic behavior
keeping in mind that it concerns only the experimentally accessible  part of
the asymptotic limit. On the other hand, in many ultrasonic experiments the
measured frequency range is limited only to one or two decades so for the
magnets with slow spin-lattice relaxation this asymptotic region can not be
approached at all. Because the ``adiabatic'' behavior comes from the $2%
\tilde{\omega}^2|b|^2g^2$ term in the coefficient $\tilde{W}_2$ we can also
say that as the reduced frequency $\tilde{\omega}$ increases, the
Murata-Iro-Schwabl behavior turns into the ``adiabatic'' one.

(ii) For $vt^{-\alpha }\Phi \ll 1$ we recover the asymptotic behavior (\ref
{asym}) but with $\tilde{g}^2$ replaced by $g^2$. What is rather unexpected
we observe the asymptotic behavior in quite a nonasymptotic region because
here $\tilde{\omega}\gg 1$ and $t$ is also large. Moreover, the
proportionality coefficient $g^2$ is exactly equal to that from the theory
which neglects all effects of the energy-density fields. However this region
is not fully critical because as a result of smallness of $vt^{-\alpha }\Phi 
$ also the critical contribution to the specific-heat is very small so the
specific-heat does not show any noticeable divergence in this region. The
last statement is easily  seen in the static case, where the temperature
dependent specific-heat is given by $C_S(t)=C_S(1+\frac v2\langle S_{{\bf k}%
}^2S_{-{\bf k}}^2\rangle _H)$, remembering that the four-spin response
function $2\langle \Gamma \widetilde{S^2}_{{\bf -k},-\omega }S_{{\bf k}%
,\omega }^2\rangle ^{L_{\text{A}}}$ transforms into $\langle S_{{\bf k}%
}^2S_{-{\bf k}}^2\rangle _H$ for $\omega =0$. Here the static average $%
\langle ...\rangle _H$ was calculated with the weight $\exp (-H)$.

In the above we have assumed that despite of $\tilde{\omega}\gg 1$ the
frequency $\bar{\omega}$ is still small. This assumption not always seems to
be correct and if so the reduced frequency $\bar{\omega}$ may become
comparable with $1$. Then the coefficients $\tilde{W}_i$ and $b(\omega )$
began to evolve, however the ``adiabatic'' limit ($\tilde{\omega}\rightarrow
\infty $) does not change as the term $2\tilde{\omega}^2|b|^2g^2$ in $\tilde{%
W}_2$ is reduced with the dominant term in the denominator of Eq.(\ref{alfa}%
).

\subsubsection{Comparison of the ``adiabatic'' limit with critical
ultrasonic attenuation \protect\\in binary mixtures and liquid $^4$He}

The analytic relation (\ref{mod}) resembles the corresponding asymptotic
formulas for sound attenuation in binary mixtures \cite
{kroll,ferrell,folk98,deng}, where the sound attenuation exponent is also
equal to $z\nu -\alpha$. Looking at Eq.(\ref{mod}) it is easy to see that it
can also be transformed into the form

\begin{equation}
\alpha (\omega ,t)\propto -g^2\omega {\rm Im}(1+v\langle \Gamma \widetilde{%
S^2}_{{\bf -k},-\omega }S_{{\bf k},\omega }^2\rangle ^{L_{\text{A}}})^{-1},
\label{adiab}
\end{equation}
in this region, which is also identical in form to the expression used by
Ferrel and Bhattacharjee \cite{ferrell80} for the ultrasonic attenuation
near the $\lambda $-transition in liquid helium (although the specific-heat
exponent is very close to zero for such systems): 
\begin{equation}
\alpha (\omega ,t)\propto -\omega {\rm Im}C_{FB}(t,\omega {)^{-1},}
\label{fb}
\end{equation}
where $C_{FB}(t,\omega )$ is a phenomenological frequency-dependent
specific-heat. Comparing Eqs.(\ref{adiab}) and (\ref{fb}) we get the
correspondence between these two equations by interpreting 
\begin{equation}
C_S^{\text{A}}(t,\omega )\equiv C_S[1+v\langle \Gamma \widetilde{S^2}_{{\bf %
-k},-\omega }S_{{\bf k},\omega }^2\rangle ^{L_{\text{A}}}]  \label{cart}
\end{equation}
as a frequency-dependent specific-heat in the Ferrel-Bhattacharjee sense,
where as usual we put $k=0$. A very similar identification, in the case of $%
^4$He, was made by Pankert and Dohm \cite{pank} who gave the statistical
meaning to $C_{FB}(t,\omega )$. Note however that in spite of the fact that
for $\omega \rightarrow 0$ Eq.( \ref{cart}) transforms (up to bilinear terms
in small couplings $w$ and $g$) into the exact relation known from the
statics 
\begin{equation}
C_S(t,k)=\bar{C}_S+(f-wg\bar{c}_0^2)^2\bar{C}_S^2\langle S_{{\bf k}}^2S_{-%
{\bf k}}^2\rangle _H,  \label{stat}
\end{equation}
where $C_S(t,k)\equiv \langle e_{{\bf k}}^Se_{-{\bf k}}^S\rangle _H$ is the
static temperature and wave vector dependent specific-heat, with $\bar{C}%
_S^{-1}=C_S^{-1}-w^2\bar{c}_0^{-2}$ and $\bar{c}_0^{-2}=c_0^{-2}-a^2w^2C_L$,
the interpretation of Eq.(\ref{adiab}) in terms of true frequency-dependent
specific-heat is inaccurate as in dynamics the  latter  is given by the
product of the spin-energy propagator $D_S(k,\omega )$ and the
frequency-dependent damping coefficient $\Gamma _S(k,\omega )$, which can be
found as the coefficient staying in front of the $\tilde{e}_S\tilde{e}_S$
term in the Lagrangian after transformation (\ref{d}). From the
transformations (\ref{a},\ref{c}-\ref{d}) we find 
\begin{equation}
D_S(k,\omega )=\tilde{\tilde{D}}_{S0}(k,\omega )+2\tilde{\tilde{D}}_{S0}^2(k,\omega )\tilde{f%
}_1(\omega )\tilde{f}_2(\omega )\langle \Gamma \widetilde{S^2}_{{\bf -k}%
,-\omega }S_{{\bf k},\omega }^2\rangle ^{L_{\text{eff}}}  \label{resprop}
\end{equation}
and $\Gamma _S(k,\omega )\simeq \Gamma _{S0}(k,\omega )$ if we drop the
terms of order of $w^2$. The determined in this way frequency-dependent
specific-heat, although giving the same static limit (\ref{stat}) for $%
\omega \rightarrow 0$ differs significantly from $C_S^{\text{A}}(t,\omega )$
for high frequencies $(\tilde{\omega}\rightarrow \infty )$
where $\tilde{\tilde{D}}_{S0}(k,\omega )\rightarrow 0$ contrary to $\bar{C}_S$ which
is a constant. Also the four-spin response functions in Eqs.(\ref{cart}) and
(\ref{resprop}) show different behavior as the coupling constant $u$ is 
frequency independent in the idealized Lagrangian $L_{\text{A}}$ in
contrast to $u(k,\omega )$ in $L_{\text{eff}}$ (Eq.(\ref{u})). The advantage
of the function $C_S^{\text{A}}(t,\omega )$ is that it is much simpler in
calculations, as $L_{\text{A}}$ does not contain the details of the
dynamics. It is worth noting that $C_S^{\text{A}}(t,\omega )$ is defined
with the help of the constant $v=2f^2C_S$ which is assumed to be finite even
if the coupling constants $g$ and $w$ are very small i.e. also in the
so-called ``weak-coupling'' limit \cite{paw89,deng}.

The ``adiabatic'' formula (\ref{adiab}) shows a close analogy to the
critical sound attenuation in two quite different systems belonging to
different universality classes and in the case of $^4$He even the order
parameter dimensionality is different, $n=2$ and $\alpha \simeq 0$. It may
suggest that as concerns the adiabatic sound propagation this kind of
singularity may be quite common in the critical systems where the order
parameter is coupled to the energy and sound mode by two different coupling
constants. The exception is the gas-liquid system but there, as was noted by
Pankert and Dohm \cite{pank}, the order-parameter itself is proportional to
the energy and there is only one static coupling between the order-parameter
and the sound variable.

\subsection{Regime II}

Now let us consider a regime which is more appropriate for description of
the common lattice consisting mostly of phonons. The new regime, opposite to
regime I, is defined by the inequality: $m\ll \bar{\omega}$ ($\omega \gg 
\frac{\gamma _L-\gamma +\lambda _{+}k^2}{C_L}$). Now the function $b(\omega
)\rightarrow \frac{m+\bar{\omega}^2+i\bar{\omega}(1-m)}{\bar{\omega}^2}$.
Let us assume further that $m\simeq \frac{\gamma _L-\gamma +\lambda _{+}k^2}{%
\Gamma _L(k)}\ll 1$ and $m\ll \bar{\omega}^2$ then $b(\omega )\simeq 1+\frac 
i{\bar{\omega}}$, $P(\omega )\simeq 1-a+ia\frac m{\bar{\omega}}$ and $-i%
\tilde{\omega}b(\omega )\simeq \frac{C_S}{C_L}(1-i\bar{\omega})$. Inserting
these expressions into Eq.(\ref{sigma3}) and neglecting the terms of order
of $\frac m{\bar{\omega}}$, or higher, we get 
\begin{equation}
\frac{vC_S\Sigma (k,\omega )}{k^2}=\frac{2\hat{g}^2C_{L0}(\omega )C_S^{\text{%
A}}(t,\omega )+2g^2C_S\Delta C_S^{\text{A}}(t,\omega )+[w^2v(1-a)^2C_S-2\hat{%
g}^2]C_SC_{L0}(\omega )}{C_S^{\text{A}}(t,\omega )+C_{L0}(\omega )},
\label{c1}
\end{equation}
where $C_{L0}(\omega )=\frac{C_L}{1-i\bar{\omega}}$ is the free,
frequency-dependent, lattice specific-heat, $\hat{g}=g-wf(1-a)C_S$ and $%
\Delta C_S^{\text{A}}(t,\omega )=C_S^{\text{A}}(t,\omega )-C_S$ is the
fluctuation only contribution to $C_S^{\text{A}}(t,\omega )$.

\subsubsection{Low-frequency and high-frequency regions}

It is seen from Eq.(\ref{c1}) that as long as $\mid C_{L0}(\omega )\mid \gg
\mid C_S^{\text{A}}(t,\omega )\mid $ i.e. when $\bar{\omega}$ is not too
high the denominator is only slightly singular and we have a kind of
competition, with frequency-dependent weights, between two kinds of
low-frequency contributions, the Murata-Iro-Schwabl and Kawasaki term. Note
only, the different amplitude, proportional to $\hat{g}^2$ in the limit $%
\tilde{\omega}=0$, instead of $\tilde{g}^2$ in Eqs.\ (\ref{asym}) and (\ref
{kawbeh}). In this low-frequency region again the spin and lattice systems
stay at equilibrium with each other. For very high $\bar{\omega}$: $\mid
C_{L0}(\omega )\mid \simeq \frac{C_L}{\bar{\omega}}\ll C_S$, which is
equivalent to the condition $\tilde{\omega}\gg 1$ the specific-heat $C_S^{%
\text{A}}(t,\omega )$ begins to outweigh in the denominator of Eq.(\ref{c1})
and the ``adiabatic'' term proportional to 
\[
g^2\frac{\Delta C_S^{\text{A}}(t,\omega )}{C_S^{\text{A}}(t,\omega )}=\text{%
const}-g^2C_S\frac 1{C_S^{\text{A}}(t,\omega )}
\]
dominates the acoustic self-energy, because the spin and lattice subsystems
are no longer in equilibrium.

\subsubsection{Relevant frequency-dependent specific-heats}

Introducing $C_{-}^{\text{A}}(t,\omega )\equiv \frac{C_{L0}(\omega )C_S^{%
\text{A}}(t,\omega )}{C_{L0}(\omega )+C_S^{\text{A}}(t,\omega )}$ and $%
C_{+}^{\text{A}}(t,\omega )\equiv C_{L0}(\omega )+C_S^{\text{A}}(t,\omega )$%
, Eq.(\ref{c1}) can also be written as 
\begin{equation}
\frac{vC_S\Sigma (k,\omega )}{k^2}=2\hat{g}^2(t,\omega )C_{-}^{\text{A}%
}(t,\omega )+\frac{2g^2C_S\Delta C_S^{\text{A}}(t,\omega )+[w^2v(1-a)^2C_S-2%
\hat{g}^2]C_SC_{L0}(\omega )}{C_{+}^{\text{A}}(t,\omega )}.  \label{c2}
\end{equation}
$C_{-}^{\text{A}}(t,\omega )$ is a kind of specific heat which by analogy
with the static, $\omega =0$, case can be given the following
interpretation. The energy contribution to the functional $H$, describing
the statics, can be rewritten as 
\[
\frac{e_S^2}{C_S}+\frac{e_L^2}{C_L}=\frac{e^2}{C_{+}}+\frac{e_\Delta ^2}{%
C_{-}}
\]
where $1/C_{-}=1/C_S+1/C_L$ and $e_\Delta =C_{-}(\frac{e_S}{C_S}-\frac{e_L}{%
C_L})$. If the lattice and the spin systems are at equilibrium with each
other, then the local fluctuations of temperature $\delta T_S=\frac{e_S}{C_S}$
and $\delta T_L=\frac{e_L}{C_L}$ are equal, and consequently $\Delta
=e_\Delta =0$. Thus $e_\Delta $ can be interpreted as excess energy in one
system (and deficiency in the other) in comparison with the case when they
were at equilibrium with each other ($\delta T_S=\delta T_L$), so if we
shift the energy $\delta E_{S\text{ }}$ from the spin system to the lattice
system ($\delta E_L=-\delta E_S$) it results in the shifts of temperatures $%
\delta T_S=\frac{\delta E_S}{C_SV}$ and $\delta T_L=\frac{\delta E_L}{C_LV}=-%
\frac{\delta E_S}{C_LV}$, where $V$ is the total volume, in both subsystems.
The total temperature difference 
\begin{equation}
T_S-T_L=\delta T_S-\delta T_L=\frac{\delta E_S}{C_SV}+\frac{\delta E_S}{C_LV}%
=\frac{\delta E_S}{C_{-}V}  \label{cmin}
\end{equation}
is proportional to the energy transported from one system to the other and $%
C_{-}V=\frac{\delta E_S}{T_S-T_L}$ can be understood as the net heat
capacity for the process of differentiation (or equalization) of
temperatures between both subsystems, whereas the total specific heat $C_{+}$
characterizes the thermal properties of the system as the whole (after
equilibrization).

Coming back to Eq.(\ref{c2}) note that it can also be rewritten as 
\begin{equation}
\frac{vC_S\Sigma (k,\omega )}{k^2}=2\hat{g}^2(t,\omega )\tilde{C}_{-}^{\text{%
A}}(t,\omega )+\frac{2(g^2-\hat{g}^2)C_S\Delta C_S^{\text{A}}(t,\omega
)+w^2v(1-a)^2C_S^2C_{L0}(\omega )}{C_{+}^{\text{A}}(t,\omega )}  \label{c3}
\end{equation}
where $\tilde{C}_{-}^{\text{A}}(t,\omega )\equiv \frac{(C_{L0}(\omega
)+C_S)\Delta C_S^{\text{A}}(t,\omega )}{C_{L0}(\omega )+C_S+\Delta C_S^{%
\text{A}}(t,\omega )}$ may be looked upon again as the net specific-heat for
the process of equilibrization for two sub-systems with the only difference
that this time the background term $C_S$ has been ``moved'' from the spin to
the lattice sub-system. Now the lattice degrees of freedom plus
short-wavelength spin fluctuations consist together the one (noncritical)
sub-system and the long-wavelength (critical) spin fluctuations the other.
In Eqs.\ (\ref{c2}) and (\ref{c3}) the second terms vanish for $g=0$ and $w=0
$, respectively. We shall call them the mixing terms because in the case
when one of the coupling constants $g$ or $wf(1-a)C_S$ dominates $\hat{g}$, 
 the acoustic self-energy is composed of the dominant term proportional to
the corresponding specific-heat $\tilde{C}_{-}^{\text{A}}(t,\omega )$ or $
C_{-}^{\text{A}}(t,\omega )$ and a small correction resulting from a
slightly different ways the constants $g$ and $w$ couple the sound to the
spin system \cite{trans}. The former one couples the sound mode only to the
part of the spin system, the long wavelength critical spin fluctuations, in
contrast to the latter coupling which couples effectively the whole spin
system (including the noncritical background) to the sound. Looking at $%
\tilde{C}_{-}^{\text{A}}(t,\omega )$ it is easy to see that for $C_L\gg C_S$
it contains both low-frequency terms: the Murata-Iro-Schwabl and Kawasaki
terms, as well as the ``adiabatic'' limit. Namely, the first term in 
\begin{equation}
{\rm Im}\tilde{C}_{-}^{\text{A}}(t,\omega )=\frac{\mid \Delta C_S^{\text{A}%
}(t,\omega )\mid ^2{\rm Im}C_{L0}(\omega )+\mid C_{L0}(\omega )+C_S\mid ^2%
{\rm Im}\Delta C_S^{\text{A}}(t,\omega )}{\mid C_{L0}(\omega )+C_S^{\text{A}%
}(t,\omega )\mid ^2}
\end{equation}
can be identified as the Kawasaki term for $\tilde{\omega}\ll 1$ ($%
C_{L0}(\omega )\gg C_S^{\text{A}}(t,\omega )$) and the second as the
Murata-Iro-Schwabl one. The crossover temperature $t_{\text{cross}}=\left( {%
\frac{\alpha ^2\gamma }{4K_4\nu ^2vC_S\Gamma }}\right) ^{\frac 1{z\nu
-\alpha }}$ depends mostly on the ratio of bare relaxation frequencies $%
\omega _S/\Gamma $. In the opposite case $\tilde{\omega}\gg 1$, $%
C_{L0}(\omega )\ll C_S^{\text{A}}(t,\omega )$, and ${\rm Im}\tilde{C}_{-}^{%
\text{A}}(t,\omega )\simeq \frac{C_S{}^2{\rm Im}\Delta C_S^{\text{A}%
}(t,\omega )}{\mid C_S^{\text{A}}(t,\omega )\mid ^2}$ i.e. the ``adiabatic''
limit is recovered. The situation described above may be also superimposed
by the effects associated with the crossover in the specific-heat $C_S^{%
\text{A}}(t,\omega )$, because the expression for $C_S^{\text{A}}(t,\omega
)=C_S(1+vt^{-\alpha }\Phi )$ may not display the required $t^{-\alpha }$
singularity in the physical region of interest (if $v$ is very small). Then,
for example, the measured effective sound attenuation exponent in the
Kawasaki region will be much smaller then $2\alpha $, sometimes close to
zero.

In closing we should mention that we have not discussed yet the third term
in the numerator of Eq.\ (\ref{alfa2}) as well as its analogue in Eq.(\ref
{c1}). It gives the weakest singularity $\propto t^{-\alpha }$ which is
usually smaller then the first two terms. However, as it is proportional, in
the $\omega =0$ limit, to $wfC_S\tilde{g}$ i.e. to the first power of $%
\tilde{g}$, whereas $W_1$ and $W_2$ are proportional to the second power,
then the  $t^{-\alpha }$ term may dominate the ultrasonic attenuation in the
degenerated case of $\tilde{g}\rightarrow 0$.

\section{SUMMARY}

We have performed a detailed analysis of the critical sound attenuation in
magnets. The relaxation of spin-energy to the lattice has been fully taken
into account in the prediction of the temperature and frequency dependencies
of the acoustic self-energy. The important point here is that the acoustic
self-energy is reducible with respect to the energy propagators. We were
able to express $\Sigma (k,\omega )$ in terms of the four-spin response
function $\langle \Gamma \widetilde{S^2}_{{\bf -k},-\omega }S_{{\bf k}%
,\omega }^2\rangle ^{L_{\text{A}}}$ of the idealized, phonon- and
energy-free model A. The last quantity can be relatively easy calculated by
the renormalization group method. In the low-frequency region, two kinds of
critical behavior described by Eqs.\ (\ref{asym}) and (\ref{kawbeh}) compete
with each other and the relative weights are determined by the coupling
constants $g$ and $w$ as well as the frequencies $\omega $, $\omega _S$, $%
\omega _L$ and the other parameters. The high-frequency or ``adiabatic''
limit is determined completely only by one coupling constant $g$ -
describing the interaction of sound mode with two spin fluctuations. The
dynamic scaling functions for these three regimes are calculated within the
one-loop approximation.

Another point of interest is the possibility of expressing  the acoustic
self-energy by a suitable frequency-dependent specific-heat. We have shown
that in the case of one coupling $g$ or $wfC_S(1-a)$ being much stronger
then the other, the acoustic self-energy can be very well approximated by
the specific-heat $C_{-}^{\text{A}}(t,\omega )$ or $\tilde{C}_{-}^{\text{A}%
}(t,\omega )$, the net specific-heats for the process of equilibrization
between the lattice and spin sub-systems or in the latter case between the
noncritical background composed of the lattice degrees of freedom plus the
noncritical short-wavelength spin fluctuations and the critical subsystem
composed of the critical long-wavelength spin fluctuations. The difference
in definitions of $C_{-}^{\text{A}}(t,\omega )$ and $\tilde{C}_{-}^{\text{A}%
}(t,\omega )$ comes from a slightly different ways the couplings $g$ and $%
wfC_S(1-a)$ couple the acoustic phonon to the fluctuations of the order
parameter. The virtue of Eqs.(\ref{c2}) and (\ref{c3}) is that (depending on
the relative strength of the coupling constants) in some regions of
parameters the acoustic self-energy can be expressed entirely by the
frequency dependent specific-heat $\tilde{C}_{-}^{\text{A}}(t,\omega )$ (or $%
C_{-}^{\text{A}}(t,\omega )$), governing the equalization of temperatures
between the lattice and spin systems. $\tilde{C}_{-}^{\text{A}}(t,\omega )$
shows its relevance in the ultrasonic propagation in magnetic systems due to
the fact that depending on the frequency, the reduced temperature and the
ratio $\omega _S/\Gamma $ it can reveal the most important types of
singularities in the critical sound attenuation.

A similar analysis is also possible for the magnets with the order parameter
dimensionality greater than one. For example the Eq.\ (\ref{sigma3}) is
still valid there. However, the specific-heat exponent $\alpha $ is negative
in X-Y and Heisenberg systems, therefore, the scaling behavior of the
Kawasaki as well as the ``adiabatic'' limit will change.

% figures follow here
% Here is an example of the general form of a figure:
% Fill in the caption in the braces of the \caption{} command. Put the label
% that you will use with \ref{} command in the braces of the \label{} command.
% \begin{figure}
% \caption{}
% \label{}
% \end{figure}
% tables follow here
% Here is an example of the general form of a table:
% Fill in the caption in the braces of the \caption{} command. Put the label
% that you will use with \ref{} command in the braces of the \label{} command.
% Insert the column specifiers (l, r, c, d, etc.) in the empty braces of the
% \begin{tabular}{} command.
% \begin{table}
% \caption{}
% \label{}
% \begin{tabular}{}
% \end{tabular}
% \end{table}

\begin{table}[tbp]
\caption{Propagators of the model}
\label{tab}
\begin{tabular}{ccc}
Propagator & Analytic expression & Free Lagrangian average \\ \hline
Spin response- & $H_0(k,\omega )=\frac 1{-i\omega +\Gamma (r+k^2)}$ & $%
\langle \widetilde{S}_{{\bf k,}\omega }S_{-{\bf k},-\omega }\rangle _0$ \\ 
Spin correlation- & $K_{S0}(k,\omega )=\frac{2\Gamma }{\omega ^2+\Gamma
(r+k^2)}$ & $\langle S_{{\bf k,}\omega }S_{-{\bf k},-\omega }\rangle _0$ \\ 
Phonon response- & $G_0(k,\omega )=\frac 1{c^2k^2+\omega ^2-i\omega Dk^2}$ & 
$\langle \widetilde{Q}_{{\bf k},\omega }Q_{-{\bf k,}-\omega }\rangle _0$ \\ 
Phonon correlation- & $K_{Q0}(k,\omega )=\frac{2Dk^2}{(c^2k^2-\omega
^2)^2-\omega ^2D^2k^4}$ & $\langle Q_{{\bf k},\omega }Q_{-{\bf k,}-\omega
}\rangle $ \\ 
Spin-energy response- & $D_{S0}(k,\omega )=\frac 1{-i\omega +\Gamma
_S(k)C_S^{-1}}$ & $\langle \tilde{e}_{{\bf k},\omega }^Se_{-{\bf k},-\omega
}^S)\rangle _0$ \\ 
Spin-energy correlation- & $K_{e_S0}(k,\omega )=\frac{2\Gamma _S(k)}{\omega
^2+\Gamma _S(k)^2}$ & $\langle e_{{\bf k},\omega }^Se_{-{\bf k},-\omega
}^S)\rangle _0$ \\ 
Lattice-energy response- & $D_{L0}(k,\omega )=\frac 1{-i\omega +\Gamma
_L(k)C_L^{-1}}$ & $\langle \tilde{e}_{{\bf k},\omega }^Le_{-{\bf k},-\omega
}^L)\rangle _0$ \\ 
Lattice-energy correlation- & $K_{e_L0}(k,\omega )=\frac{2\Gamma _L(k)}{%
\omega ^2+\Gamma _L(k)^2}$ & $\langle e_{{\bf k},\omega }^Le_{-{\bf k}%
,-\omega }^L)\rangle _0$%
\end{tabular}
\end{table}


\begin{references}
\bibitem{trans}  The mixing term in Eq.(\ref{c2}) participates also in
forming the correct ``adiabatic'' limit which always is proportional to $g^2$%
.
\end{references}
\end{document}